\documentclass[reprint,amsmath,amssymb,aps,pra]{revtex4-2}
\usepackage{braket}
\usepackage{graphicx}
\usepackage{dcolumn}
\usepackage{bm}
\usepackage{booktabs}

\usepackage{color}
\usepackage{xcolor}
\usepackage{hyperref}

\newcommand{\rhoone}{\rho^{(1)}}                 
\newcommand{\rhonet}{\rho^{(1)}_{\mathrm{NN}}}      
\newcommand{\K}{\mathsf{K}}                       
\newcommand{\F}{\mathsf{F}}                       
\newcommand{\Dm}{\mathsf{D}}                      

\begin{document}

\title{PINNverse meets Jordan--Wigner: learning the Hamiltonian of long spin
chains}

\author{Gubio G. de Lima}%
\email{gubiofisika@gmail.com}
\affiliation{Physics Department, Federal University of S\~ao Carlos,
13565-905, S\~ao Carlos, SP, Brazil}

\date{\today}

\begin{abstract}
Inverse physics-informed neural networks (PINNverse) learn the parameters of a
quantum Hamiltonian from measured expectation values, but the network output
grows as $4^N$ with the number of qubits $N$. We show that, for spin chains
with nearest-neighbor exchange interactions, the Jordan-Wigner transformation
removes this exponential cost: the dynamics closes on an $N\times N$ one-body
density matrix, so the network needs only $N^2$ outputs and the data are local
observables measured in five global settings. Parasitic $ZZ$ couplings
(crosstalk) are included through a time-dependent Hartree--Fock closure. With
synthetic data, the same network first solves the equation of motion and then
recovers the Hamiltonian parameters of chains of up to $50$ qubits, with
couplings learned to better than $0.1\%$. Errors in the local frequencies of
long chains are removed by a direct least-squares fit started from the network
estimate.
\end{abstract}

\maketitle

\section{Introduction}\label{sec:intro}

Physics-informed neural networks (PINNs) are neural networks trained to
represent the solution of a differential equation: besides (or instead of)
labeled data, the loss function contains the residual of the governing
equations evaluated at a set of collocation points, with derivatives computed
by automatic differentiation~\cite{RAISSI2019686,Karniadakis2021,
doi:10.1137/19M1274067,doi:10.1126/sciadv.abi8605}. The same construction
solves inverse problems: when the coefficients of the equations are unknown,
they are promoted to trainable parameters and learned together with the
solution from a small set of measurements. Following
Refs.~\cite{Castelano2024,deLima2025}, we refer to this inverse formulation as
PINNverse.

Determining the Hamiltonian that generates the dynamics of a quantum device is
a central task in the calibration of quantum hardware and in quantum
simulation~\cite{Wiebe2014,Gebhart2023,Bairey2019,Li2020,Wittler2021}.
Traditional approaches such as process tomography~\cite{Chuang1997} require a
number of measurement settings that grows exponentially with the number of
qubits. PINNverse was shown to learn general two-qubit Hamiltonians from the
time evolution of a few expectation values, and, combined with a freezing
mechanism that isolates pairs of qubits, to address larger
registers~\cite{Castelano2024}. It was subsequently extended to open quantum
systems governed by Lindblad master equations, where it identifies both
Hamiltonian parameters and decay rates and detects crosstalk
couplings~\cite{deLima2025}. Related physics-informed approaches have been
applied to quantum control~\cite{Norambuena2024} and to Hamiltonian learning
with neural representations of the quantum state~\cite{LiuWang2025}.

In all these formulations the network represents an object whose size is
exponential in the number of qubits $N$: either the density matrix $\rho(t)$,
with $4^N$ real entries, or the complete set of Pauli expectation values
$\langle\sigma_{\mu_1}\otimes\cdots\otimes\sigma_{\mu_N}\rangle(t)$, with
$4^N-1$ entries (15 for the two-qubit problems of
Refs.~\cite{Castelano2024,deLima2025}). The physical loss requires the
commutator with the Hamiltonian, whose cost grows even faster. For $N=20$ the
network would need about $10^{12}$ outputs. Representing the state with a
neural-network quantum state~\cite{Carleo2017} reduces the number of network
parameters, but the physical loss still involves the $2^N$ amplitudes, and
inverse-PINN demonstrations of this type have been limited to about ten
spins~\cite{LiuWang2025}. The freezing
mechanism~\cite{Castelano2024} circumvents this by learning the device two
qubits at a time, at the price of additional control and of many independent
experiments. A formulation that learns a whole chain at once, with a number of
network outputs that grows polynomially with $N$, is still missing.

Spin-$1/2$ chains are not directly described by canonical bosonic or
fermionic operators, which makes the analytical treatment of many-body
Hamiltonians nontrivial. In one spatial dimension, however, the Jordan--Wigner
transformation~\cite{Jordan1928,Lieb1961,Coleman2015,Landi2019} provides an
exact mapping between spin operators and fermionic operators by encoding the
nonlocal commutation structure into a string operator. This strategy is
particularly useful because it converts the transverse spin-exchange terms
into quadratic hopping terms, thereby simplifying the Hamiltonian and, in
special cases, reducing the problem to a non-interacting fermionic model. The
key observation of this work is that, for such quadratic models, the
Heisenberg equations of motion close on the $N\times N$ one-body density matrix
of the fermions. The exponential object that the network had to represent is
then replaced by a matrix of $N^2$ real numbers, without any approximation.

This class of Hamiltonians is not an academic restriction: chains of
superconducting transmon qubits coupled by exchange interactions implement
precisely the inhomogeneous $XX$ model in a longitudinal field, and they are
routinely used as analog quantum simulators~\cite{Roushan2017,Yan2019,
Hangleiter2024}. The dominant coherent error in these devices is a residual
$ZZ$ interaction between neighboring qubits~\cite{Krantz2019,Chen2014,Sung2021,
Mi2021}, which is the typical crosstalk mechanism addressed in
Refs.~\cite{Sarovar2020,deLima2025}. Under the Jordan--Wigner mapping, $ZZ$
crosstalk becomes a weak density--density interaction between fermions, which
we treat with a time-dependent Hartree--Fock closure.

\begin{figure*}[t]
\centering
\includegraphics[width=\textwidth]{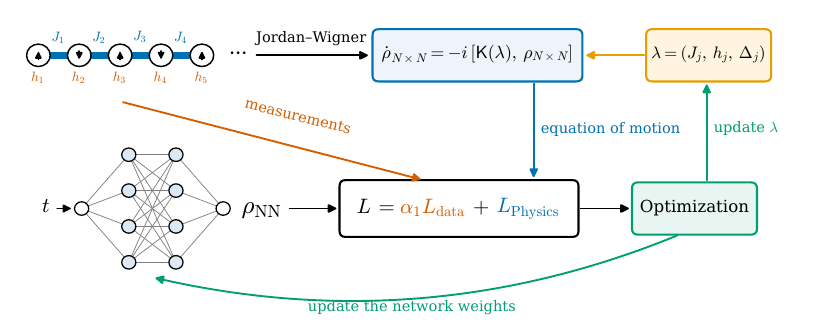}
\caption{Workflow of the method. Top: a chain of $N$ qubits with unknown
parameters $\bm\lambda=(J_j,h_j,\Delta_j)$ is mapped by the Jordan--Wigner
transformation onto an equation of motion for the $N\times N$ one-body
density matrix $\rho_{N\times N}\equiv\rhoone(t)$, which replaces the $4^N-1$
Pauli expectation values. Bottom: a neural network receives only the time $t$
and returns $\rho_{\mathrm{NN}}\equiv\rhonet(t)$. In the loss of
Eq.~\eqref{eq:Loss_total}, the data term compares $\rho_{\mathrm{NN}}$ with
local measurements of the chain and the physics term is the residual of the
equation of motion. The optimization updates the network weights and
$\bm\lambda$ together.}
\label{fig:scheme}
\end{figure*}

The method is summarized in Fig.~\ref{fig:scheme}. Our contributions are: (i) a reformulation of PINNverse for spin chains in which the network represents the fermionic one-body density matrix, reducing
the number of outputs from $4^N-1$ to $N^2$ while keeping the dynamics exact
for the inhomogeneous $XX$ chain; (ii) the identification of a minimal set of
local observables, accessible with five global measurement settings for any
$N$, that makes all hopping amplitudes and field differences identifiable;
(iii) a Hartree--Fock extension for $ZZ$ crosstalk with an error of second
order in the crosstalk strength; and (iv) a numerical demonstration in which the network first solves the
equation of motion (forward problem) and then learns the parameters (inverse
problem) for chains of up to $N=50$ qubits, including noisy data and
crosstalk detection, with a direct least-squares fit as a classical baseline. Section~\ref{sec:model_methods} presents the model, the
Jordan--Wigner mapping, the resulting equations of motion and the PINNverse
procedure. Section~\ref{sec:applications} contains the applications.
Section~\ref{sec:discussion} discusses limitations and extensions, including
open systems, and Sec.~\ref{sec:conclusions} concludes. Derivations and
implementation details are given in the Supplemental Material.

\section{Theoretical model and methodology}\label{sec:model_methods}

\subsection{Model Hamiltonian}\label{subsec:model}

We consider an open (non-periodic) chain of $N$ qubits described by
\begin{equation}
\begin{split}
    \hat{H}(\bm{\lambda})
    ={}&
    \sum_{j=1}^{N-1}\frac{J_j}{2}
    \left(\hat\sigma^x_j\hat\sigma^x_{j+1}+\hat\sigma^y_j\hat\sigma^y_{j+1}\right)
    +\sum_{j=1}^{N}\frac{h_j}{2}\,\hat\sigma^z_j
    \\
    &+\sum_{j=1}^{N-1}\Delta_j\,\hat\sigma^z_j\hat\sigma^z_{j+1},
\end{split}
\label{eq:H_param}
\end{equation}
where $\hbar=1$, $J_j$ are the exchange (hopping) couplings, $h_j$ are the
local qubit frequencies (detunings), and $\Delta_j$ are $ZZ$ couplings. In the
notation of Refs.~\cite{Castelano2024,deLima2025}, the Hamiltonian is a linear
combination $\hat H(\bm\lambda)=\sum_m\lambda_m\hat G_m$ of the generators
$\hat G_m\in\{\tfrac12(\hat\sigma^x_j\hat\sigma^x_{j+1}+\hat\sigma^y_j
\hat\sigma^y_{j+1}),\ \tfrac12\hat\sigma^z_j,\ \hat\sigma^z_j
\hat\sigma^z_{j+1}\}$ with parameters
$\bm\lambda=(J_1,\dots,J_{N-1},h_1,\dots,h_N,\Delta_1,\dots,\Delta_{N-1})$.

We study two applications. In the first, $\Delta_j=0$: Eq.~\eqref{eq:H_param}
is the inhomogeneous $XX$ chain in a longitudinal field, which is the
Hamiltonian implemented by chains of exchange-coupled transmon
qubits~\cite{Roushan2017,Yan2019,Hangleiter2024}; the goal is to calibrate all
couplings and frequencies. In the second, small nonzero $\Delta_j$ model the
residual $ZZ$ crosstalk present in these devices~\cite{Krantz2019,Sung2021,
Mi2021}; the goal is to detect and quantify it. The homogeneous anisotropic
Heisenberg chain of Refs.~\cite{Coleman2015,Landi2019},
$\hat H=-J\sum_j(\hat S^x_j\hat S^x_{j+1}+\hat S^y_j\hat S^y_{j+1})
-J_z\sum_j\hat S^z_j\hat S^z_{j+1}$ with $\hat S^\alpha=\hat\sigma^\alpha/2$,
is recovered with $J_j=-J/2$, $\Delta_j=-J_z/4$ and $h_j=0$
(Supplemental Material, Sec.~S1).

\subsection{Jordan--Wigner transformation}\label{subsec:jw}

We start from the local identification between the spin basis and the
occupation basis of a fermionic mode,
\begin{equation}
    |\uparrow\rangle \equiv f^\dagger |0\rangle,
    \qquad
    |\downarrow\rangle \equiv |0\rangle,
\end{equation}
from which the on-site spin operators can be written as
\begin{equation}
    \hat\sigma^{+}=f^\dagger,\qquad
    \hat\sigma^{-}=f,\qquad
    \hat\sigma^{z}=2f^\dagger f-1,
\end{equation}
with $\hat\sigma^\pm=(\hat\sigma^x\pm i\hat\sigma^y)/2=\hat S^\pm$. For a
single site, this representation is sufficient. However, for a spin chain,
independent spin operators commute whereas fermionic operators anticommute. To
restore the correct algebra in one dimension, we introduce the Jordan--Wigner
string operator and define~\cite{Jordan1928,Coleman2015}
\begin{equation}
    \hat\sigma_j^{+}=f_j^\dagger e^{i\phi_j},\qquad
    \hat\sigma_j^{-}=f_j e^{-i\phi_j},\qquad
    \hat\sigma_j^{z}=2n_j-1,
    \label{eq:jw}
\end{equation}
with
\begin{equation}
    \phi_j=\pi\sum_{l<j} n_l,
    \qquad
    n_l=f_l^\dagger f_l,
\end{equation}
and $\{f_k,f_l^\dagger\}=\delta_{kl}$. Since $e^{i\pi n_l}=1-2n_l=-\hat
\sigma^z_l$, the string is $e^{i\phi_j}=\prod_{l<j}(-\hat\sigma^z_l)$. The
nonlocal phase factor $e^{i\phi_j}$ guarantees that the mapped operators
satisfy the appropriate spin commutation relations along the chain.

\paragraph{Logic of the simplification.}
The central idea of the method is that the string operator transfers the
statistical information of all sites to the left of site $j$ into a phase
factor. As a consequence, the spin problem is rewritten in terms of fermionic
occupation and hopping processes. For nearest neighbors the strings cancel:
using $e^{i\phi_j}e^{-i\phi_{j+1}}=e^{-i\pi n_j}$ and $f_j^\dagger(1-2n_j)=
f_j^\dagger$,
\begin{equation}
    \hat\sigma^+_j\hat\sigma^-_{j+1}
    = f_j^\dagger e^{-i\pi n_j} f_{j+1}
    = f_j^\dagger f_{j+1}.
\end{equation}
Since $\hat\sigma^x_j\hat\sigma^x_{j+1}+\hat\sigma^y_j\hat\sigma^y_{j+1}
=2(\hat\sigma^+_j\hat\sigma^-_{j+1}+\hat\sigma^-_j\hat\sigma^+_{j+1})$, the
transverse spin interaction is mapped into a fermionic hopping term. Likewise,
the longitudinal terms become
\begin{equation}
    \hat\sigma^z_j\hat\sigma^z_{j+1}=4n_jn_{j+1}-2n_j-2n_{j+1}+1 .
\end{equation}
Up to an additive constant, the fermionized Hamiltonian assumes the form
\begin{equation}
\begin{split}
    \hat H
    ={}&
    \sum_{j=1}^{N-1}J_j\left(f_j^\dagger f_{j+1}+f_{j+1}^\dagger f_j\right)
    +\sum_{j=1}^{N}h_j n_j
    \\
    &+\sum_{j=1}^{N-1}\Delta_j\left(4n_jn_{j+1}-2n_j-2n_{j+1}\right).
\end{split}
    \label{eq:H_jw_fermionic}
\end{equation}
Equation~\eqref{eq:H_jw_fermionic} makes explicit the gain of the
transformation: the original spin Hamiltonian is recast into a fermionic model
composed of hopping, density, and nearest-neighbor interaction terms. For
$\Delta_j=0$ the interaction vanishes and the problem becomes exactly
quadratic~\cite{Lieb1961},
\begin{equation}
    \hat H_{0}=\sum_{k,l=1}^{N}\K_{kl}\,f_k^\dagger f_l,
    \label{eq:H_quadratic}
\end{equation}
with $\K_{jj}=h_j$ and $\K_{j,j+1}=\K_{j+1,j}=J_j$,
where $\K(\bm\lambda)$ is a real, symmetric, tridiagonal $N\times N$ matrix.

Two remarks are important for what follows. First, we work with an open
chain. With periodic boundary conditions the bond between sites $N$ and $1$
carries the string of the whole chain and becomes
$J_N f_N^\dagger f_1(-1)^{\hat N_F}$, where $\hat N_F=\sum_jn_j$; the
Hamiltonian then splits into even- and odd-parity sectors with different
boundary conditions~\cite{Landi2019}. This complication is absent in the open
chain, which is also the geometry of most devices. Second, because the
parameters $J_j$ and $h_j$ are inhomogeneous, translation invariance is
broken and the momentum-space diagonalization usually employed for these
models (Supplemental Material, Sec.~S1) is not available. We therefore work
directly in real space.

\subsection{Time-dependent quantum dynamics}\label{subsec:dynamics}

For a closed quantum system, the density matrix $\rho(t)$ evolves according to
the Liouville--von Neumann equation~\cite{Breuer2002}:
\begin{equation}
    \frac{d\rho(t)}{dt} = -\frac{i}{\hbar}\,[\hat{H}(\bm\lambda), \rho(t)].
    \label{eq:lvn_t}
\end{equation}
For $N$ qubits, $\rho(t)$ is a $2^N\times2^N$ matrix. Equivalently, the
Heisenberg equations for the $4^N-1$ Pauli expectation values form a closed
linear system~\cite{deLima2025}. Either way, the number of dynamical variables
is exponential in $N$.

\paragraph{One-body density matrix.}
For the quadratic Hamiltonian~\eqref{eq:H_quadratic} a much smaller set of
variables is sufficient. We define the one-body density matrix of the
Jordan--Wigner fermions,
\begin{equation}
    \rhoone_{lk}(t)=\langle f_k^\dagger f_l\rangle(t)
    =\mathrm{Tr}\big[\rho(t)\,f_k^\dagger f_l\big],
    \label{eq:rho1_def}
\end{equation}
a Hermitian, positive semidefinite $N\times N$ matrix with eigenvalues in
$[0,1]$ and trace equal to the number of excitations
$\hat N_F=\sum_j(1+\hat\sigma^z_j)/2$. Using the identity
$[f_a^\dagger f_b,f_k^\dagger f_l]=\delta_{bk}f_a^\dagger f_l-
\delta_{al}f_k^\dagger f_b$ in the Heisenberg equation
$d\hat O/dt=i[\hat H,\hat O]$, we obtain (Supplemental Material, Sec.~S2)
\begin{equation}
    \frac{d\rhoone}{dt}=-i\left[\K(\bm\lambda),\rhoone\right],
    \quad
    \rhoone(t)=e^{-i\K t}\rhoone(0)\,e^{i\K t}.
    \label{eq:lvn_rho1}
\end{equation}
Equation~\eqref{eq:lvn_rho1} has exactly the form of the Liouville--von Neumann
equation~\eqref{eq:lvn_t}, but for an $N\times N$ matrix generated by the
single-particle Hamiltonian $\K$. It is exact for \emph{any} initial state,
because the commutator of a quadratic Hamiltonian with a bilinear is again a
bilinear: the hierarchy of equations of motion closes at the level of two-point
functions. Evaluating its right-hand side costs $\mathcal{O}(N^3)$ operations
for dense matrices, or $\mathcal{O}(N^2)$ exploiting the tridiagonal
structure of $\K$, instead of the $\mathcal{O}(8^N)$ required by
Eq.~\eqref{eq:lvn_t}.

\paragraph{Measured observables.}
The data used to train the network are expectation values of local spin
operators that are linear functionals of $\rhoone$. Using
Eqs.~\eqref{eq:jw} and~\eqref{eq:rho1_def},
\begin{subequations}\label{eq:observables}
\begin{align}
    \langle\hat\sigma^z_j\rangle&=2\rhoone_{jj}-1,\\
    \langle\hat\sigma^x_j\hat\sigma^x_{j+1}+\hat\sigma^y_j\hat\sigma^y_{j+1}
    \rangle&=4\,\mathrm{Re}\,\rhoone_{j,j+1},\\
    \langle\hat\sigma^x_j\hat\sigma^y_{j+1}-\hat\sigma^y_j\hat\sigma^x_{j+1}
    \rangle&=4\,\mathrm{Im}\,\rhoone_{j,j+1}.
\end{align}
\end{subequations}
The last quantity is the spin current through bond $(j,j+1)$. These $3N-2$
time series involve only one-qubit and nearest-neighbor two-qubit Pauli
operators, and they can be acquired with five global measurement settings,
independently of $N$: all qubits in $Z$; all in $X$; all in $Y$; $X$ on odd
and $Y$ on even sites; and $Y$ on odd and $X$ on even sites. For comparison,
reconstructing all Pauli expectation values requires $3^N$ settings.

\paragraph{Initial state.}
We use the N\'eel state $|{\uparrow\downarrow\uparrow\downarrow\cdots}\rangle$,
prepared from the ground state with single-qubit $\pi$ pulses, for which
$\rhoone(0)=\mathrm{diag}(1,0,1,0,\dots)$. Being a Slater determinant, it is
a Gaussian (quasi-free) fermionic state, fully characterized by
$\rhoone$~\cite{Peschel2003}, and it remains Gaussian under the evolution
generated by $\hat H_0$. Equation~\eqref{eq:lvn_rho1} also holds for the
product state $|+\rangle^{\otimes N}$ used in Refs.~\cite{Castelano2024,
deLima2025}, for which $\rhoone(0)=\tfrac12\mathbb{1}$ plus $\tfrac14$ on
the first off-diagonals. That state, however, is not Gaussian and contains
coherences between sectors of different fermion parity; its single-qubit
coherences $\langle\hat\sigma^x_j\rangle$ involve strings and cannot be
expressed through $\rhoone$, and the Hartree--Fock closure introduced below is
not justified for it. The N\'eel state is therefore the natural choice.

\paragraph{ZZ crosstalk: Hartree--Fock closure.}
For $\Delta_j\neq0$ the quartic term $4\Delta_jn_jn_{j+1}$ in
Eq.~\eqref{eq:H_jw_fermionic} couples the equation for $\rhoone$ to four-point
functions, and the hierarchy no longer closes. We close it at the
Gaussian level using Wick's theorem,
$\langle n_an_b\rangle\simeq\rhoone_{aa}\rhoone_{bb}-|\rhoone_{ab}|^2$,
which yields the time-dependent Hartree--Fock (TDHF)
equation~\cite{Ring1980} (Supplemental Material, Sec.~S3)
\begin{equation}
    \frac{d\rhoone}{dt}=-i\left[\F\big(\bm\lambda,\rhoone\big),\rhoone\right],
    \label{eq:tdhf}
\end{equation}
with the self-consistent Fock matrix
\begin{subequations}\label{eq:fock}
\begin{align}
    \F_{jj}&=h_j+2\Delta_{j-1}\langle\hat\sigma^z_{j-1}\rangle
            +2\Delta_j\langle\hat\sigma^z_{j+1}\rangle,\\
    \F_{j,j+1}&=J_j-4\Delta_j\,\rhoone_{j,j+1},
    \qquad \F_{j+1,j}=\F_{j,j+1}^{*},
\end{align}
\end{subequations}
where $\langle\hat\sigma^z_m\rangle=2\rhoone_{mm}-1$ and
$\Delta_0=\Delta_N=0$. The diagonal term is the Hartree (mean-field)
contribution: each qubit feels a frequency shift proportional to the
magnetization of its neighbors, which is the familiar picture of $ZZ$
crosstalk as a state-dependent frequency shift. The off-diagonal term is the
Fock (exchange) correction to the hopping. In compact form,
$\F=\K+\mathrm{diag}(2\Dm\bm z)-4\,\Dm\circ\rhoone$, where $\Dm$ is the
tridiagonal matrix of couplings $\Delta_j$ with zero diagonal, $z_m=
\langle\hat\sigma^z_m\rangle$, and $\circ$ is the element-wise product.
Equation~\eqref{eq:tdhf} is nonlinear in $\rhoone$, has no closed-form
solution, preserves $\mathrm{Tr}\,\rhoone$, Hermiticity and the Hartree--Fock
energy, and reduces to the exact Eq.~\eqref{eq:lvn_rho1} for $\Delta_j=0$.

\paragraph{Regime of validity.}
For a Gaussian initial state the closure is exact to first order in $\Delta$:
the error in $d\rhoone/dt$ is proportional to $\Delta$ times the
non-Gaussianity of the state, and the non-Gaussianity generated during the
evolution is itself of order $\Delta$. The error in the observables at a fixed
time is therefore $\mathcal{O}(\Delta^2)$, whereas neglecting the crosstalk
altogether produces an error $\mathcal{O}(\Delta)$. The approximation is
expected to be accurate for $|\Delta_j|/|J_j|\ll1$ and moderate times, which is
precisely the crosstalk regime of superconducting chains: for transmons with
anharmonicity $U$ the residual $ZZ$ coupling scales as $J^2/|U|$, typically
one to two orders of magnitude below the exchange coupling, and it can be
further suppressed with tunable couplers~\cite{Krantz2019,Sung2021,Mi2021}. For $\Delta\sim J$ (Heisenberg
regime) tensor-network methods are required~\cite{Vidal2004,Wilde2022}.
A numerical check of this scaling against the exact spin dynamics is
given in Sec.~\ref{subsec:validation}.

\subsection{Physics-informed neural networks for the inverse problem}
\label{sec:pinn_inverse}

A neural network is a composition of parametrized functions built from
artificial neurons organized in layers~\cite{nielsen2015Neural,Roberts2022}.
For a specific neuron $j$ at layer $l$, the output is
\begin{equation}
    y_j^l(\bm{\theta} , a) = \sigma \!\left( \sum_{k} w_{jk}^l a_k^{l-1} + b_j^l
    \right),
\end{equation}
where $w_{jk}^l$ and $b_j^l$ are the trainable weights and biases,
$a_k^{l-1}$ is the input or the output of the previous layer, and
$\sigma(\cdot)$ is a nonlinear activation function. The set of network
parameters is denoted by $\bm{\theta}=\{w_{jk}^l,b_j^l\}$. The PINN method
extends how neural networks learn by adding new cost functions to their
training, so that the network approximates the solution of the governing
differential equations. In the forward problem the differential equation is
fully known; in our case this means knowing that we solve
Eq.~\eqref{eq:lvn_rho1} [or Eq.~\eqref{eq:tdhf}] and knowing the Hamiltonian,
i.e., the generators $\hat G_m$ and the parameters $\bm\lambda$. In the
inverse problem the generators are known but $\bm\lambda$ is not, and
$\bm\lambda$ is promoted to a set of trainable parameters.

\paragraph{Network output and exact initial condition.}
The network receives the time $t$, normalized to $s=2t/T-1\in[-1,1]$, and returns
a vector $\bm y(t)$ of exactly $N^2$ real numbers: the $N(N+1)/2$ entries of the
upper triangle of a real symmetric matrix $Y_R$ and the $N(N-1)/2$ entries of
the strict upper triangle of a real antisymmetric matrix $Y_I$. The one-body
density matrix is constructed as
\begin{equation}
    \rhonet(t)=\rhoone(0)+\frac{t}{T}\left[Y_R(t)+i\,Y_I(t)\right],
    \label{eq:net_output}
\end{equation}
where $T$ is the final time and $\rhoone(0)$ is the prepared initial state.
This ansatz is Hermitian by construction, has no redundant outputs, and
satisfies the initial condition exactly for any value of the network
parameters (a hard constraint~\cite{Lu2021hard}), so that no initial-condition
term has to be balanced against the other terms of the loss. In the original
PINNverse the network returns $4^N-1$ Pauli expectation values~\cite{deLima2025};
here it returns $N^2$ numbers.

We describe the loss as
\begin{equation}\label{eq:Loss_total}
    L = \alpha_1 L_{\mathrm{data}} + L_{\mathrm{Physics }},
\end{equation}
where $ L_{\mathrm{Physics }} = \alpha_2 L_{\mathrm{eq}} + \alpha_3
L_{\mathrm{others}}$ and the weights $\alpha_k$ balance each contribution. To describe each term of the cost function, we
define $t_j$ as the times at which the equation of motion is enforced and
$t_n$ as the times of the experimental data.

\paragraph{Data term.} We assume that the experimental data are always
expectation values. Thus,
\begin{equation}
    L_{\mathrm{data}}  = \frac{1}{N_c}\sum_{n=1}^{N_c}\sum_{\mu=1}^{3N-2}
    \Big| \langle \hat{O}_\mu \rangle^{\mathrm{exp}}(t_n) - \langle
    \hat{O}_\mu^{\mathrm{NN}}\rangle(t_n) \Big|^2,
    \label{eq:L_data}
\end{equation}
where $\langle \hat{O}_\mu \rangle^{\mathrm{exp}}(t_n)=\mathrm{Tr}[\rho(t_n)
\hat{O}_{\mu}]$ is the measured value of the observable $\hat O_\mu$ at time
$t_n$, with $\hat O_\mu$ running over the $3N-2$ local observables of
Eq.~\eqref{eq:observables}, and $N_c$ is the number of collocation points with
data. The prediction $\langle\hat{O}_\mu^{\mathrm{NN}}\rangle(t_n)$ is
obtained from the network output $\rhonet(t_n)$ through the linear relations
of Eq.~\eqref{eq:observables}. When only the data loss is used, we obtain a
supervised learning scenario. This term is essential for our model because it
restricts the behavior of $\rhonet$ at each instant of time; it is, however, a
penalty on the network output only. Note that there is still no restriction
on the parameters we want to find, $\bm{\lambda}$, as they are not included in
this term, and it would be impossible to find them using only this term. Note
also that the data constrain only $3N-2$ of the $N^2$ entries of $\rhonet$;
the remaining entries are fixed by the equation of motion and the initial
condition.

The term $L_{\mathrm{Physics}}$ can have several contributions that depend on
the physical system. Here it contains the equation of motion and a
conservation law that plays the role of the trace condition of the density
matrix; the initial condition is already built into
Eq.~\eqref{eq:net_output}.

\paragraph{Differential equation term.} It enforces Eq.~\eqref{eq:lvn_rho1}
[or Eq.~\eqref{eq:tdhf}] at the collocation points $\{t_j\}_{j=1}^{N_t}$:
\begin{equation}
    L_{\mathrm{eq}} = \frac{1}{N_t}\sum_{j=1}^{N_t}
    \Bigg\|
        \frac{d\rhonet}{dt}
        + i \Big[\F\big(\bm{\lambda},\rhonet\big),\rhonet\Big]
    \Bigg\|^2_{t=t_j},
    \label{eq:L_eq_heis}
\end{equation}
where $\|\cdot\|$ is the Frobenius norm and $\F(\bm\lambda,\rhoone)=
\K(\bm\lambda)$ in the exact case $\Delta_j=0$. In this way, the network's
solution $\rhonet(t_j)$ is penalized if it is not the correct evolution for
the current set of parameters $\bm{\lambda}$. This is similar to solving
differential equations (forward problem) with the PINN
method~\cite{RAISSI2019686,doi:10.1126/sciadv.abi8605,doi:10.1137/19M1274067}.
Now the parameters $\bm{\lambda}$ enter a constraint: while
Eq.~\eqref{eq:L_data} drives the network towards the correct one-body density
matrix, the loss~\eqref{eq:L_eq_heis} evaluated on this correct matrix
vanishes only if $\bm{\lambda}$ is correct. In the initial training phase both
are wrong; therefore, the data loss plays a more important role than the
equation-of-motion loss. The time derivatives of the $N^2$ network outputs
are obtained by automatic differentiation. Because the input is a scalar, all
of them are computed with a single forward-mode (tangent) pass through the
network, instead of one reverse pass per output
(Supplemental Material, Sec.~S5). The collocation points $t_j$ are a fixed uniform
grid plus a set of random points that is redrawn at every training
step~\cite{Wu2023}.

\paragraph{Physical regularizers.} The analog of the normalization condition
$\mathrm{Tr}[\rho(t)]=1$ is the conservation of the number of excitations,
$\mathrm{Tr}[\rhoone(t)]=N_{\mathrm{exc}}$, which holds because
Eq.~\eqref{eq:H_param} commutes with the total magnetization
$\sum_j\hat\sigma^z_j$. We impose it through
\begin{equation}\label{eq:L_norm}
    L_{\mathrm{number}}=\frac{1}{N_t}\sum_{j=1}^{N_t}
    \Big|\mathrm{Tr}\big[\rhonet(t_j)\big]-N_{\mathrm{exc}}\Big|^2,
\end{equation}
so that $L_{\mathrm{others}}=L_{\mathrm{number}}$, with $N_{\mathrm{exc}}=
\lceil N/2\rceil$ for the N\'eel state. This condition is implied by
Eq.~\eqref{eq:lvn_rho1}; the explicit term is the analog of the trace term
used for the full density matrix in Ref.~\cite{deLima2025}.

\paragraph{Forward and inverse problems.} The same network and the same loss
solve two problems. In the \emph{forward} problem $\bm\lambda$ is known,
$\alpha_1=0$, and the network is trained to solve the equation of motion; no
data are used. In the \emph{inverse} problem $\bm\lambda$ is trainable,
initialized at the nominal chain ($J_j=\bar J$, $h_j=0$, $\Delta_j=0$), and
the data term is switched on. The true parameters never enter the loss nor
the choice of hyperparameters; they are used only to generate the synthetic
data and to score the result.

\paragraph{Optimization.} Training has two stages. The first is
Adam~\cite{Kingma2015} with a cosine learning-rate schedule and a larger
learning rate for $\bm\lambda$ than for the network weights. First-order
optimizers, however, stall on a plateau where the fitted trajectory already
looks correct while the parameters, which enter the loss only through the
time derivatives, are still inaccurate~\cite{Rathore2024}. The second stage is
therefore the quasi-Newton method L-BFGS~\cite{LiuNocedal1989} with a
strong-Wolfe line search on a fixed dense collocation grid, in double
precision. All quantities are computed in double precision from the start,
because the line search of the second stage is sensitive to round-off
errors.
\begin{table*}[t]
    \caption{Size of the learning problem as a function of the number of qubits
    $N$. Pauli PINNverse: network output with all Pauli expectation
    values~\cite{deLima2025}. JW-PINNverse: one-body density matrix
    [Eq.~\eqref{eq:net_output}]. Settings: global measurement settings required
    by the data loss. Parameters: unknowns of Eq.~\eqref{eq:H_param} with $h_1=0$.}
    \label{tab:complexity}
    \begin{ruledtabular}
    \begin{tabular}{lccccc}
     & $N=2$ & $N=10$ & $N=20$ & $N=50$ & $N=100$\\
    \hline
    Pauli outputs, $4^N-1$ & $15$ & $1.0\times10^{6}$ & $1.1\times10^{12}$ &
    $1.3\times10^{30}$ & $1.6\times10^{60}$\\
    JW outputs, $N^2$ & $4$ & $100$ & $400$ & $2500$ & $10^{4}$\\
    Pauli settings, $3^N$ & $9$ & $5.9\times10^{4}$ & $3.5\times10^{9}$ &
    $7.2\times10^{23}$ & $5.2\times10^{47}$\\
    JW settings & $5$ & $5$ & $5$ & $5$ & $5$\\
    Parameters, $3N-3$ & $3$ & $27$ & $57$ & $147$ & $297$\\
    \end{tabular}
    \end{ruledtabular}
\end{table*}
\subsection{Identifiability and gauge symmetries}\label{subsec:identifiability}

Not every parameter in Eq.~\eqref{eq:H_param} can be learned from the
dynamics of $\rhoone$, and the reduced set of observables introduces discrete
ambiguities that must be resolved by the choice of data.

\paragraph{Uniform field.} A uniform shift $h_j\to h_j+c$ adds $c\hat N_F$ to
the Hamiltonian, which commutes with $\rhoone$ for any state of fixed
magnetization, so it leaves the dynamics invariant. Only the differences
$h_j-h_1$ are identifiable; we fix $h_1=0$ and learn the remaining $N-1$
fields. The number of trainable physical parameters is $2N-2$ for the $XX$
chain and $3N-3$ with crosstalk.

\paragraph{Discrete symmetries.} For a real initial matrix $\rhoone(0)$, such
as the N\'eel state, the local gauge transformation $f_j\to s_jf_j$ with
$s_j=\pm1$ maps $J_j\to s_js_{j+1}J_j$ and $\rhoone_{kl}\to s_ks_l\rhoone_{kl}$,
while the time-reversal (complex conjugation) symmetry maps
$\bm\lambda\to-\bm\lambda$ and $\rhoone\to\rhoone{}^{*}$. The diagonal
$\rhoone_{jj}$, i.e., the magnetizations $\langle\hat\sigma^z_j\rangle$, is
invariant under both. Therefore, using only $\langle\hat\sigma^z_j\rangle(t)$,
the signs of the individual couplings and the global sign of $(\bm J,\bm h)$
cannot be determined. The nearest-neighbor correlations
$\mathrm{Re}\,\rhoone_{j,j+1}$ change sign under the gauge transformation that
flips $J_j$, and the currents $\mathrm{Im}\,\rhoone_{j,j+1}$ change sign under
time reversal; including both sets in $L_{\mathrm{data}}$ removes all
ambiguities. With crosstalk, the TDHF equation~\eqref{eq:tdhf} keeps both
symmetries, with the gauge transformation leaving $\Delta_j$ unchanged and
time reversal flipping all parameters, $(\bm J,\bm h,\bm\Delta)\to
-(\bm J,\bm h,\bm\Delta)$. Once the currents fix the global sign of
$(\bm J,\bm h)$, the signs of the $\Delta_j$ are therefore also determined.

\subsection{Computational complexity}\label{subsec:complexity}

Table~\ref{tab:complexity} compares the size of the problem in the original
PINNverse formulation and in the present one. The network output, the number
of measurement settings and the cost of the physical loss become polynomial in
$N$. The number of unknown parameters grows linearly, as it does in any local
Hamiltonian.

\section{Applications}\label{sec:applications}

\paragraph{Protocol.} Synthetic data are generated by drawing
$J_j\in[0.5,1.5]\bar J$ and $|h_j|\in[0.3,1.0]\bar J$ with random signs
($h_1=0$) from uniform distributions; time is measured in units of
$1/\bar J$. For the $XX$ chain the exact data are obtained from
Eq.~\eqref{eq:lvn_rho1} by diagonalizing $\K$, which is feasible for
arbitrary $N$. For the crosstalk problem the data are generated from the exact
spin dynamics of Eq.~\eqref{eq:H_param} (state-vector simulation) for $N=8$,
so that the error of the Hartree--Fock closure is included in the assessment.
Unless stated otherwise we use $t\in[0,4/\bar J]$, $N_t=200$ fixed plus $100$
random collocation points for $L_{\mathrm{eq}}$, and $N_c=50$ equally spaced
data points. Measurement noise is modeled, as in Ref.~\cite{deLima2025}, by
adding to each data point a random number drawn from a normal distribution
$\mathcal N(0,\sigma^2)$. The accuracy of the estimated parameters is
quantified by the mean absolute percentage error
\begin{equation}
    \mathrm{MAPE}=\frac{1}{D}\sum_{i=1}^{D}\frac{|P_i^{\mathrm{exact}}-
    P_i^{\mathrm{pred}}|}{|P_i^{\mathrm{exact}}|},
    \label{eq:mape}
\end{equation}
where $D$ is the number of parameters of a given type and
$P_i^{\mathrm{exact}}$ ($P_i^{\mathrm{pred}}$) is the exact (predicted)
value. Because a small data misfit does not by itself guarantee correct
parameters, we also report the root-mean-square deviation between the exact
observables and those re-simulated with the learned parameters. Every
inverse problem is also solved, on the same data and from the same initial
guess, by a direct least-squares fit that uses no neural network
(Supplemental Material, Sec.~S5). The network architecture, the optimizer
settings and the initial guess are the same in all applications
(Supplemental Material, Sec.~S5), and each configuration is trained once, with a
single random seed.

\subsection{Validation of the mapping}\label{subsec:validation}

\begin{figure}[t]
\centering
\includegraphics[width=\columnwidth]{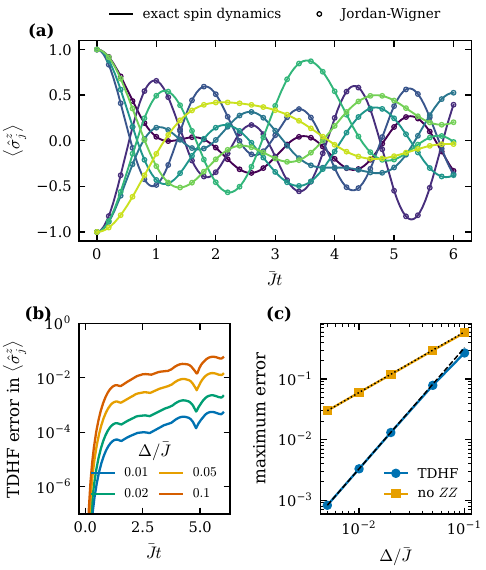}
\caption{Validation of the Jordan--Wigner description for an $N=8$ chain
prepared in the N\'eel state. (a) Magnetization $\langle\hat\sigma^z_j\rangle(t)$
of all qubits obtained from the exact spin dynamics (lines) and from
Eq.~\eqref{eq:lvn_rho1} (symbols) for $\Delta_j=0$. (b) Site-averaged error of
$\langle\hat\sigma^z_j\rangle(t)$ in the TDHF approximation,
Eq.~\eqref{eq:tdhf}, for several values of $\Delta/\bar J$. (c) Maximum error
of the observables~\eqref{eq:observables} over $t\in[0,6/\bar J]$ as a
function of $\Delta/\bar J$ for TDHF (circles) and for the model that neglects
the $ZZ$ coupling (squares); the dashed and dotted lines indicate slopes $2$
and $1$.}
\label{fig:validation}
\end{figure}

Before training, we verify the mapping against the exact spin dynamics. For
$\Delta_j=0$, the observables~\eqref{eq:observables} computed from
Eq.~\eqref{eq:lvn_rho1} coincide with those of the full state-vector
evolution to machine precision (maximum deviation $8\times10^{-15}$ for $N=8$,
Fig.~\ref{fig:validation}(a)), confirming the signs of
Eqs.~\eqref{eq:H_jw_fermionic}--\eqref{eq:observables}. For $\Delta_j\neq0$
we quantify the error of the TDHF closure as a function of $\Delta/\bar J$
and of time, and compare it with the error of neglecting the crosstalk
[Fig.~\ref{fig:validation}(b,c)]. For uniform $\Delta_j=\Delta$ and
$t\le6/\bar J$, the maximum error of the TDHF observables is $3.3\times10^{-3}$,
$1.3\times10^{-2}$ and $7.9\times10^{-2}$ for $\Delta/\bar J=0.01$, $0.02$ and $0.05$, which
follows the expected quadratic scaling, whereas neglecting the $ZZ$ term
gives $6.0\times10^{-2}$, $1.2\times10^{-1}$ and $3.0\times10^{-1}$, which grows linearly.

\subsection{Forward problem: solving the equation of motion}
\label{subsec:forward}

\begin{figure*}[t]
\centering
\includegraphics[width=\textwidth]{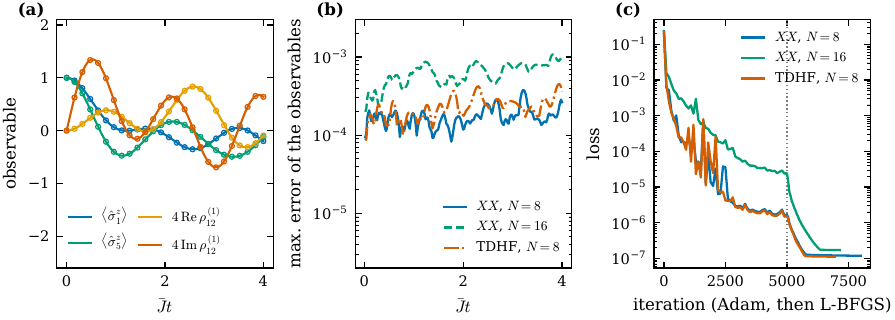}
\caption{Forward problem: the network solves the equation of motion with known
parameters and no data. (a) Reference solution (lines) and network (symbols)
for representative observables of the $XX$ chain with $N=8$. (b) Maximum
error of the observables~\eqref{eq:observables} as a function of time for the
$XX$ chain with $N=8$ and $N=16$ and for the TDHF equation with $N=8$ and
$\Delta/\bar J=0.05$. (c) Loss during training; the first $5000$ iterations
(left of the dotted line) are Adam and the remaining ones L-BFGS.}
\label{fig:forward}
\end{figure*}

We first show that the network can solve the differential equation itself.
With the parameters $\bm\lambda$ fixed at their exact values and without any
data ($\alpha_1=0$), the network is trained only on
$L_{\mathrm{eq}}+L_{\mathrm{number}}$, with the initial condition imposed by
Eq.~\eqref{eq:net_output}. This is the usual forward PINN and it isolates the
question of whether a single network with $N^2$ outputs can represent the
dynamics of $\rhoone$ from the question of whether the data identify the
parameters. Figure~\ref{fig:forward} compares the result with the reference
solution [closed form of Eq.~\eqref{eq:lvn_rho1} for the $XX$ chain, numerical
integration of Eq.~\eqref{eq:tdhf} with crosstalk]. The maximum deviation of
any entry of $\rhonet(t)$ from the reference over the whole time window is
$1.0\times10^{-4}$ for the $XX$ chain with $N=8$, $5.9\times10^{-4}$ for
$N=16$, and $1.4\times10^{-4}$ for the nonlinear TDHF equation with $N=8$ and
$\Delta/\bar J=0.05$. Figure~\ref{fig:forward}(c) shows the role of the two
training stages: after the Adam stage, L-BFGS lowers the loss by a further
one to two orders of magnitude, down to $1.2\times10^{-7}$ ($N=8$) and
$1.7\times10^{-7}$ ($N=16$), where it reaches a plateau.

\subsection{Inhomogeneous XX chain: calibration of couplings and frequencies}
\label{subsec:xx}

\begin{figure*}[t]
\centering
\includegraphics[width=\textwidth]{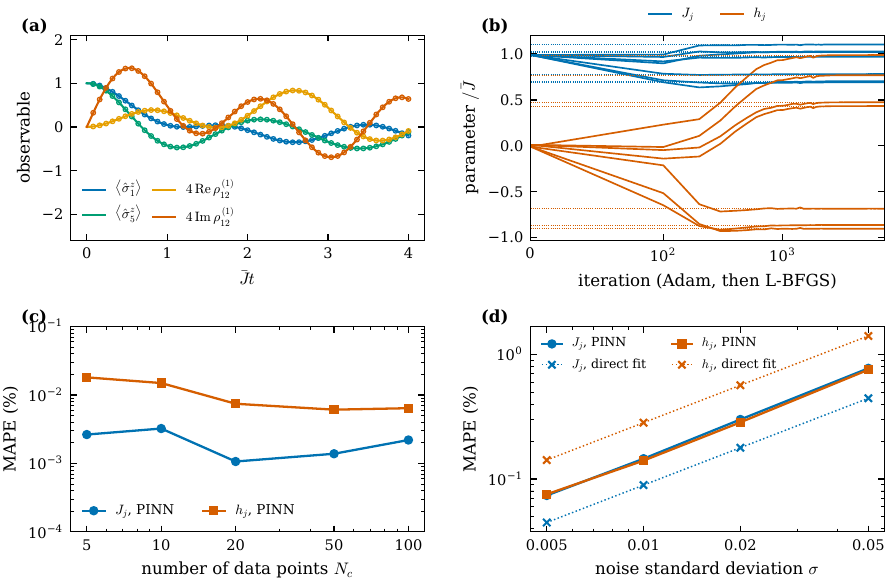}
\caption{Learning an inhomogeneous $XX$ chain with $N=8$ qubits.
(a) Measured observables~\eqref{eq:observables} (symbols, $N_c=50$ points) and
network prediction (lines) for representative sites and bonds. (b) Couplings
$J_j$ and fields $h_j$ during training (solid lines) and exact values (dotted
lines); all parameters start from $J_j=\bar J$ and $h_j=0$. (c) MAPE of $J_j$
and $h_j$ as a function of the number of data points $N_c$ for noise-free
data. (d) MAPE as a function of the standard deviation $\sigma$ of the
Gaussian noise for $N_c=50$; crosses are the direct least-squares fit without
a neural network on the same data (for the noise-free data of panel (c) the
direct fit is exact to machine precision and is not shown).}
\label{fig:xx}
\end{figure*}

Having established that the network solves the equation of motion, we turn
to the inverse problem: the calibration of all $N-1$ couplings $J_j$ and
$N-1$ field differences $h_j-h_1$ of an $XX$ chain from the $3N-2$ local time
series. Figure~\ref{fig:xx}(a,b) shows the learned dynamics and the evolution
of the parameters during training for $N=8$. For noise-free data the MAPE is
0.0014\% for the couplings and 0.0061\% for the fields, and the
observables re-simulated with the learned parameters deviate from the exact
ones by $4.9\times10^{-5}$ (root mean square). The dependence on the number of
data points $N_c$ and on the noise level $\sigma$ is shown in
Fig.~\ref{fig:xx}(c,d), in parallel with the analysis of
Ref.~\cite{deLima2025} for two qubits. With only $N_c=5$ data points the MAPE
is 0.0026\% ($J_j$) and 0.018\% ($h_j$). With noise of
standard deviation $\sigma=0.01$ the MAPE is 0.15\% ($J_j$) and
0.14\% ($h_j$), to be compared with 0.089\% and
0.28\% for the direct fit to the same data; for $\sigma=0.05$ it
is 0.78\% and 0.76\% (direct fit:
0.45\% and 1.4\%). Three observations follow. First, the accuracy depends only weakly on
$N_c$: the equation of motion interpolates between the data points, so a few
measured times are sufficient. Second, the MAPE grows linearly with $\sigma$
and is of the same order for the network and for the direct fit; in this
noise realization the network is more accurate for the fields and less
accurate for the couplings. With noise, the accuracy is therefore limited by
the information contained in the data and not by the method. Third, for
noise-free data the direct fit of the closed-form solution reaches machine
precision, whereas the network is limited by its approximation error of the
dynamics (the plateau of the loss in Fig.~\ref{fig:forward}(c)).

\subsection{Scaling with the number of qubits}\label{subsec:scaling}

\begin{figure*}[t]
\centering
\includegraphics[width=\textwidth]{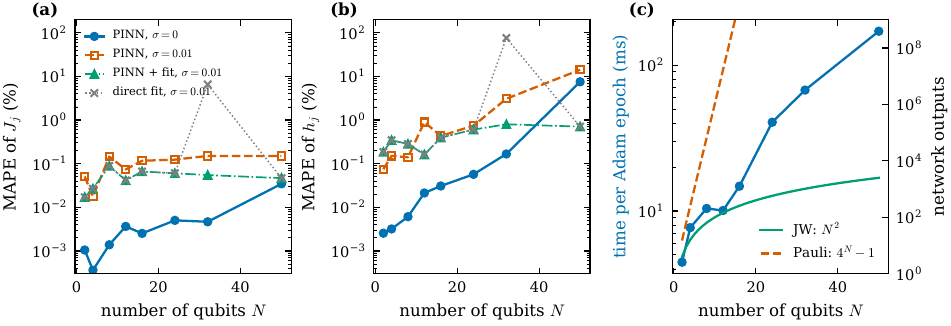}
\caption{Scaling with the chain length. MAPE of (a) the couplings $J_j$ and
(b) the fields $h_j$ as a function of $N$: network alone for noise-free data
and for $\sigma=0.01$, direct fit started from the nominal chain, and direct
fit started from the network estimate (``PINN + fit''), both for
$\sigma=0.01$. (c) Training time per Adam epoch (symbols, left axis) and
number of network outputs (lines, right axis) for the JW-PINNverse ($N^2$)
and for a network that returns all Pauli expectation values ($4^N-1$).}
\label{fig:scaling}
\end{figure*}

Because the network output grows as $N^2$, the same procedure, with the same
hyperparameters, can be applied to chains that are far beyond the reach of
the original formulation. Figure~\ref{fig:scaling} and
Table~\ref{tab:scaling} show the MAPE and the training time per epoch as a
function of $N$ up to $N=50$, for which the network has 2500
outputs, whereas the Pauli formulation would need $4^{50}-1$. One Adam
epoch for $N=50$ takes 171~ms and the whole training
91~min on a single CPU core. Since the Hamiltonian and the observables
are local, every parameter influences the data in its neighborhood from the
earliest times, and the time window does not have to grow with $N$.

The couplings are learned accurately for all chain lengths: for noise-free
data the MAPE of $J_j$ stays below $0.1\%$ up to $N=50$
(0.034\%). The fields behave differently. For noise-free data their MAPE
is below $0.2\%$ up to $N=32$, but it reaches 7.5\% for $N=50$,
and with noise $\sigma=0.01$ it is already at the percent level for $N=32$
(Table~\ref{tab:scaling}). This error is not local: the learned fields
differ from the exact ones by an offset that is nearly constant over long
segments of the chain. The reason is the gauge freedom of
Sec.~\ref{subsec:identifiability}: the local data determine the differences
$h_{j+1}-h_j$, and $h_j-h_1$ is the sum of $j-1$ of them, so that small
errors made close to the first site are carried along the whole chain. The
root-mean-square error of the local differences for $N=50$ is $2.7\times10^{-3}$
$\bar J$, much smaller than the largest error of $h_j$, $8.0\times10^{-2}$$\bar J$. In
addition, the loss is an average over $N^2$ matrix entries and $3N-2$ time
series, and the optimizer stops at a similar value of the loss for all $N$,
which corresponds to larger local errors in longer chains. The length of the
first training stage matters here: with the $5000$ Adam epochs used for the
short chains, the MAPE of the fields for noise-free data was $3.6\%$ for
$N=24$ and $17\%$ for $N=50$, against $0.057\%$ and 7.5\% with the
$20000$ epochs used for $N>16$.

The direct fit, started from the same nominal chain, converges to the
correct parameters for all chain lengths except $N=32$, where it
is trapped in a local minimum (Table~\ref{tab:scaling}). The two approaches
are complementary: when the parameters learned by the network are used as
the starting point of the direct fit (``PINN + fit''), the fit converges for
all $N$, including $N=32$, and removes the accumulated error of the
fields. For $N=50$ and $\sigma=0.01$ the MAPE is then
0.047\% for $J_j$ and 0.71\% for $h_j$, compared
with 0.15\% and 14\% for the network alone. These
results are for a single random chain and a single training run per length.

\begin{table*}[t]
\caption{Scaling with the chain length: number of network outputs, time per
Adam epoch (single CPU core), and MAPE (\%) of the couplings and fields for
the JW-PINNverse, for the direct fit started from the nominal chain, and for
the direct fit started from the network estimate (PINN + fit).}
\label{tab:scaling}
\centering
\begin{tabular}{rrrcccccccc}
\toprule
 & & & \multicolumn{2}{c}{PINN, $\sigma=0$} & \multicolumn{2}{c}{PINN, $\sigma=0.01$} & \multicolumn{2}{c}{direct fit, $\sigma=0.01$} & \multicolumn{2}{c}{PINN + fit, $\sigma=0.01$}\\
$N$ & outputs & ms/epoch & $J_j$ & $h_j$ & $J_j$ & $h_j$ & $J_j$ & $h_j$ & $J_j$ & $h_j$\\
\midrule
2 & 4 & 4.5 & 0.0011 & 0.0026 & 0.051 & 0.075 & 0.017 & 0.19 & 0.017 & 0.19\\
4 & 16 & 7.7 & $3.6\times10^{-4}$ & 0.0032 & 0.018 & 0.15 & 0.027 & 0.34 & 0.027 & 0.34\\
8 & 64 & 10.4 & 0.0014 & 0.0061 & 0.15 & 0.14 & 0.089 & 0.28 & 0.089 & 0.28\\
12 & 144 & 10.1 & 0.0037 & 0.021 & 0.074 & 0.90 & 0.041 & 0.16 & 0.041 & 0.16\\
16 & 256 & 14.8 & 0.0025 & 0.031 & 0.12 & 0.43 & 0.066 & 0.39 & 0.066 & 0.39\\
24 & 576 & 40.6 & 0.0050 & 0.057 & 0.12 & 0.74 & 0.060 & 0.61 & 0.060 & 0.61\\
32 & 1024 & 67.5 & 0.0047 & 0.17 & 0.15 & 3.1 & 6.6 & 76 & 0.055 & 0.80\\
50 & 2500 & 171.0 & 0.034 & 7.5 & 0.15 & 14 & 0.047 & 0.71 & 0.047 & 0.71\\
\bottomrule
\end{tabular}
\end{table*}

\subsection{Detection of ZZ crosstalk}\label{subsec:crosstalk}

\begin{figure*}[t]
\centering
\includegraphics[width=\textwidth]{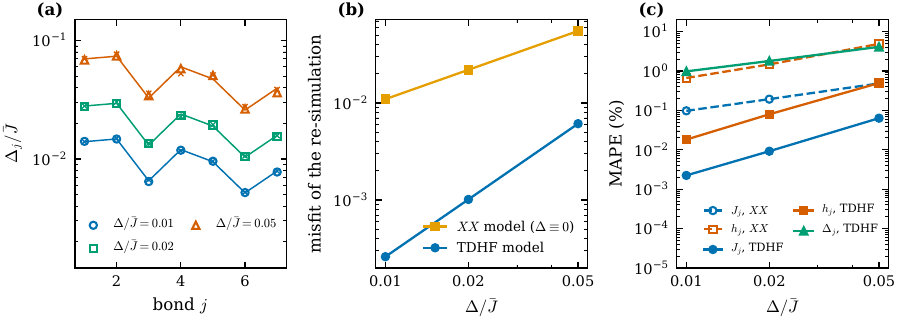}
\caption{Detection of $ZZ$ crosstalk in a chain of $N=8$ qubits with
data generated from the exact spin dynamics. (a) Learned crosstalk couplings
$\Delta_j$ (open symbols: PINN; crosses: direct fit) compared with the exact
values (lines). (b) Root-mean-square deviation between the data and the
observables re-simulated with the learned parameters, as a function of the
crosstalk strength, for the model that neglects $ZZ$ couplings and for the
TDHF model. (c) MAPE of the couplings $J_j$ and fields $h_j$ obtained with
each model, and of $\Delta_j$ with the TDHF model.}
\label{fig:crosstalk}
\end{figure*}

The second application is the detection of residual $ZZ$ couplings, the
many-qubit version of the crosstalk problem studied in Ref.~\cite{deLima2025}.
The data are generated from the exact spin dynamics of an $N=8$ chain with
$\Delta_j\in[0.5,1.5]\Delta$ and $\Delta/\bar J=0.01$, $0.02$ and $0.05$, and
the physical loss uses the TDHF equation~\eqref{eq:tdhf} with $\Delta_j$ as
additional trainable parameters, initialized at zero. We compare two models:
one that neglects crosstalk ($\Delta_j\equiv0$, exact $XX$ equations) and one
that includes it through TDHF (Fig.~\ref{fig:crosstalk} and
Table~\ref{tab:crosstalk}). Crosstalk is detected when the $XX$ model cannot reproduce the data. Because
the network is only softly constrained by the equation of motion, the fit of
the network itself to the data is not a sensitive test; we use instead the
root-mean-square deviation between the data and the observables re-simulated
with the learned parameters. It is $1.1\times10^{-2}$, $2.2\times10^{-2}$ and
$5.5\times10^{-2}$ for the $XX$ model at the three crosstalk strengths, against
$2.6\times10^{-4}$, $1.0\times10^{-3}$ and $6.1\times10^{-3}$ for the TDHF model. The MAPE of the
learned $\Delta_j$ is 0.97\%, 1.8\% and 4.1\%. The data
contain no noise, so this error is the bias of the Hartree--Fock closure; it
grows approximately linearly with $\Delta$, as expected from an error of the
observables of second order in $\Delta$. Neglecting the crosstalk also biases
the other parameters: the MAPE of the fields $h_j$ is 0.66\%,
1.5\% and 4.9\% with the $XX$ model and 0.018\%,
0.079\% and 0.49\% with the TDHF model. In these three cases
the network estimates of $\Delta_j$ and $h_j$ are closer to the exact values
than those of the direct fit of the TDHF equation (1.2\%,
2.4\% and 6.2\% for $\Delta_j$; 0.057\%, 0.22\%
and 1.3\% for $h_j$), which enforces the approximate equation
exactly.

\begin{table*}[t]
\caption{Detection of $ZZ$ crosstalk ($N=8$, noise-free data from the exact
spin dynamics): root-mean-square misfit of the observables re-simulated with
the learned parameters, and MAPE (\%) of the learned parameters, for the model without crosstalk and for the TDHF model, and MAPE
of $\Delta_j$ for the direct fit of the TDHF equation.}
\label{tab:crosstalk}
\centering
\begin{tabular}{lcccccccc}
\toprule
 & \multicolumn{3}{c}{$XX$ model ($\Delta_j\equiv0$)} & \multicolumn{4}{c}{TDHF model} & direct fit\\
$\Delta/\bar J$ & misfit & $J_j$ & $h_j$ & misfit & $J_j$ & $h_j$ & $\Delta_j$ & $\Delta_j$\\
\midrule
0.01 & $1.1\times10^{-2}$ & 0.097 & 0.66 & $2.6\times10^{-4}$ & 0.0022 & 0.018 & 0.97 & 1.2\\
0.02 & $2.2\times10^{-2}$ & 0.19 & 1.5 & $1.0\times10^{-3}$ & 0.0092 & 0.079 & 1.8 & 2.4\\
0.05 & $5.5\times10^{-2}$ & 0.48 & 4.9 & $6.1\times10^{-3}$ & 0.063 & 0.49 & 4.1 & 6.2\\
\bottomrule
\end{tabular}
\end{table*}

\section{Discussion and extensions}\label{sec:discussion}

\paragraph{Scope.} The exact reduction relies on three properties of
Eq.~\eqref{eq:H_param}: the chain is one-dimensional, the exchange couplings
connect only nearest neighbors, and there are no transverse local fields.
Next-nearest-neighbor exchange leaves a string $f_j^\dagger(1-2n_{j+1})
f_{j+2}$ that generates quartic terms; transverse fields $\hat\sigma^x_j$ are
odd fermionic operators with a string and break the quadratic structure; in
two dimensions the strings are nonlocal. Weak next-nearest-neighbor couplings,
present in real devices, can be treated with the same Hartree--Fock closure as
the $ZZ$ terms. The general two-body Hamiltonians of
Refs.~\cite{Castelano2024,deLima2025} lie outside this class, and for them the
freezing mechanism remains the appropriate strategy.

\paragraph{Anisotropic XY chain.} Anisotropic couplings
$\frac{J_j}{2}[(1+\gamma_j)\hat\sigma^x_j\hat\sigma^x_{j+1}+(1-\gamma_j)
\hat\sigma^y_j\hat\sigma^y_{j+1}]$, which include the transverse-field Ising
model, generate pairing terms $f_j^\dagger f_{j+1}^\dagger$ that do not
conserve $\hat N_F$. The model remains quadratic, and the dynamics closes on
the $2N\times2N$ Majorana covariance matrix (Supplemental Material, Sec.~S4),
with $N(2N-1)$ real entries. The procedure of Sec.~\ref{sec:pinn_inverse}
applies with this matrix as the network output.

\paragraph{Open systems.} The extension of Ref.~\cite{deLima2025} to Lindblad
dynamics carries over to the present formulation for the most relevant noise
channels. For pure dephasing, $\hat L_j=\sqrt{\gamma_j}\,\hat\sigma^z_j$, the
adjoint Lindblad equation closes exactly on $\rhoone$~\cite{Zunkovic2014}:
\begin{equation}
    \frac{d\rhoone_{lk}}{dt}=-i\left[\K,\rhoone\right]_{lk}
    -2(\gamma_k+\gamma_l)\,(1-\delta_{kl})\,\rhoone_{lk}.
\end{equation}
For amplitude damping, $\hat L_j=\sqrt{\gamma_j}\,\hat\sigma^-_j$, one
obtains $-\gamma_k\rhoone_{kk}$ on the diagonal and
$-\frac12(\gamma_k+\gamma_l)\rhoone_{lk}$ off the diagonal, plus the term
$-2\gamma_j\langle f_k^\dagger f_l n_j\rangle$ for every site $j$ strictly
between $k$ and $l$, which originates from the string of $\hat\sigma^-_j$.
This term vanishes for losses at the boundary qubits, where the problem is
quadratic and exactly solvable~\cite{Prosen2008}, and in the bulk it can be
closed with Wick's theorem as in Eq.~\eqref{eq:tdhf}. Learning the decay
rates of a chain therefore becomes possible with $\mathcal{O}(N^2)$ network
outputs.

\paragraph{Role of the neural network.} For the exact $XX$ chain,
Eq.~\eqref{eq:lvn_rho1} has a closed-form solution, and the parameters could
also be obtained by fitting $e^{-i\K t}\rhoone(0)e^{i\K t}$ directly to the
data with gradient-based optimization. The PINN formulation becomes essential
when no closed-form solution is available: in the nonlinear TDHF equation, in
open systems, and for time-dependent controls $\bm\lambda(t)$, which the
network can represent as additional outputs, as in
Ref.~\cite{Castelano2024}. For this reason every inverse problem of
Sec.~\ref{sec:applications} was also solved by such a direct fit, started
from the same nominal chain. The comparison is instructive. When it
converges, the direct fit is faster (seconds to minutes) and, for the exact
$XX$ model, more accurate than the network, which is limited by its
approximation error. Its convergence from the nominal chain is, however, not
guaranteed (it failed for $N=32$), and it requires solving the
equation of motion at every iteration. The network was trained successfully
from the nominal chain in all cases, and its estimate is a reliable starting
point for the direct fit. With an approximate model (TDHF), the soft
enforcement of the equation gave slightly less biased parameters than the
exact enforcement.

\section{Conclusion}\label{sec:conclusions}

We have shown that the Jordan--Wigner transformation removes the exponential
scaling of PINNverse for one-dimensional spin chains with nearest-neighbor
exchange. For the inhomogeneous $XX$ chain in a longitudinal field the
dynamics closes exactly on the $N\times N$ one-body density matrix of the
Jordan--Wigner fermions, which obeys a Liouville--von Neumann equation of
dimension $N$. The network therefore represents $N^2$ real numbers instead of
$4^N-1$ Pauli expectation values, the physical loss costs $\mathcal{O}(N^3)$
per time point, and all identifiable parameters are learned from one- and
two-qubit observables measured in five global settings. Uniform fields are
not identifiable and discrete gauge and time-reversal ambiguities are removed
by including nearest-neighbor correlations and spin currents in the data.
Parasitic $ZZ$ couplings are incorporated through a time-dependent
Hartree--Fock closure with error $\mathcal{O}(\Delta^2)$, appropriate for the
weak crosstalk of superconducting chains. Numerically, the network solves the equation of motion as a forward problem
and then recovers the couplings of chains of up to $N=50$ qubits with a
MAPE below $0.1\%$ from noise-free data. The errors of the local frequencies
accumulate along long chains and are removed by a direct least-squares fit
started from the network estimate, which gives 0.047\% and
0.71\% for $N=50$ with measurement noise $\sigma=0.01$.
The method also detects and quantifies $ZZ$ crosstalk from data generated
with the exact spin dynamics.
Natural next steps are the anisotropic $XY$ chain through the Majorana
covariance matrix, open chains with dephasing and boundary losses, and the
application to experimental data from superconducting quantum simulators.


\bibliography{references}

@article{RAISSI2019686,
  author  = {Raissi, M. and Perdikaris, P. and Karniadakis, G. E.},
  title   = {Physics-informed neural networks: A deep learning framework for solving forward and inverse problems involving nonlinear partial differential equations},
  journal = {J. Comput. Phys.},
  volume  = {378},
  pages   = {686},
  year    = {2019}
}

@article{doi:10.1126/sciadv.abi8605,
  author  = {Wang, S. and Wang, H. and Perdikaris, P.},
  title   = {Learning the solution operator of parametric partial differential equations with physics-informed {DeepONets}},
  journal = {Sci. Adv.},
  volume  = {7},
  pages   = {eabi8605},
  year    = {2021}
}

@article{doi:10.1137/19M1274067,
  author  = {Lu, L. and Meng, X. and Mao, Z. and Karniadakis, G. E.},
  title   = {{DeepXDE}: A deep learning library for solving differential equations},
  journal = {SIAM Rev.},
  volume  = {63},
  pages   = {208},
  year    = {2021}
}

@article{Karniadakis2021,
  author  = {Karniadakis, G. E. and Kevrekidis, I. G. and Lu, L. and Perdikaris, P. and Wang, S. and Yang, L.},
  title   = {Physics-informed machine learning},
  journal = {Nat. Rev. Phys.},
  volume  = {3},
  pages   = {422},
  year    = {2021}
}

@book{nielsen2015Neural,
  author    = {Nielsen, M. A.},
  title     = {Neural Networks and Deep Learning},
  publisher = {Determination Press},
  year      = {2015}
}

@book{Roberts2022,
  author    = {Roberts, D. A. and Yaida, S. and Hanin, B.},
  title     = {The Principles of Deep Learning Theory},
  publisher = {Cambridge University Press},
  address   = {Cambridge},
  year      = {2022}
}

@article{Carleo2017,
  author  = {Carleo, G. and Troyer, M.},
  title   = {Solving the quantum many-body problem with artificial neural networks},
  journal = {Science},
  volume  = {355},
  pages   = {602},
  year    = {2017}
}

@article{Castelano2024,
  author  = {Castelano, L. K. and Cunha, I. and Luiz, F. S. and de Jesus Napolitano, R. and Prado, M. V. d. S. and Fanchini, F. F.},
  title   = {Combining physics-informed neural networks with the freezing mechanism for general {H}amiltonian learning},
  journal = {Phys. Rev. A},
  volume  = {110},
  pages   = {032607},
  year    = {2024}
}

@article{deLima2025,
  author  = {de Lima, G. G. and Cunha, I. and Castelano, L. K.},
  title   = {Inverse physics-informed neural networks procedure for detecting noise in open quantum systems},
  journal = {arXiv:2507.12552},
  year    = {2025}
}

@article{Norambuena2024,
  author  = {Norambuena, A. and Mattheakis, M. and Gonz{\'a}lez, F. J. and Coto, R.},
  title   = {Physics-informed neural networks for quantum control},
  journal = {Phys. Rev. Lett.},
  volume  = {132},
  pages   = {010801},
  year    = {2024}
}

@article{LiuWang2025,
  author  = {Liu, J. and Wang, X.},
  title   = {Hamiltonian learning via inverse physics-informed neural networks},
  journal = {Phys. Rev. Res.},
  note    = {doi:10.1103/nx97-zjdf; arXiv:2506.10379},
  year    = {2025}
}

@article{Gebhart2023,
  author  = {Gebhart, V. and Santagati, R. and Gentile, A. A. and Gauger, E. M. and Craig, D. and Ares, N. and Banchi, L. and Marquardt, F. and Pezz{\`e}, L. and Bonato, C.},
  title   = {Learning quantum systems},
  journal = {Nat. Rev. Phys.},
  volume  = {5},
  pages   = {141},
  year    = {2023}
}

@article{Wiebe2014,
  author  = {Wiebe, N. and Granade, C. and Ferrie, C. and Cory, D. G.},
  title   = {Hamiltonian learning and certification using quantum resources},
  journal = {Phys. Rev. Lett.},
  volume  = {112},
  pages   = {190501},
  year    = {2014}
}

@article{Bairey2019,
  author  = {Bairey, E. and Arad, I. and Lindner, N. H.},
  title   = {Learning a local {H}amiltonian from local measurements},
  journal = {Phys. Rev. Lett.},
  volume  = {122},
  pages   = {020504},
  year    = {2019}
}

@article{Li2020,
  author  = {Li, Z. and Zou, L. and Hsieh, T. H.},
  title   = {Hamiltonian tomography via quantum quench},
  journal = {Phys. Rev. Lett.},
  volume  = {124},
  pages   = {160502},
  year    = {2020}
}

@article{Wilde2022,
  author  = {Wilde, F. and Kshetrimayum, A. and Roth, I. and Hangleiter, D. and Sweke, R. and Eisert, J.},
  title   = {Scalably learning quantum many-body {H}amiltonians from dynamical data},
  journal = {Quantum Sci. Technol.},
  note    = {doi:10.1088/2058-9565/ae6fe3; arXiv:2209.14328},
  year    = {2026}
}

@article{Hangleiter2024,
  author  = {Hangleiter, D. and Roth, I. and Fuksa, J. and Eisert, J. and Roushan, P.},
  title   = {Robustly learning the {H}amiltonian dynamics of a superconducting quantum processor},
  journal = {Nat. Commun.},
  volume  = {15},
  pages   = {9595},
  year    = {2024}
}

@article{Chuang1997,
  author  = {Chuang, I. L. and Nielsen, M. A.},
  title   = {Prescription for experimental determination of the dynamics of a quantum black box},
  journal = {J. Mod. Opt.},
  volume  = {44},
  pages   = {2455},
  year    = {1997}
}

@article{Wittler2021,
  author  = {Wittler, N. and Roy, F. and Pack, K. and Werninghaus, M. and Roy, A. S. and Egger, D. J. and Filipp, S. and Wilhelm, F. K. and Machnes, S.},
  title   = {Integrated tool set for control, calibration, and characterization of quantum devices applied to superconducting qubits},
  journal = {Phys. Rev. Appl.},
  volume  = {15},
  pages   = {034080},
  year    = {2021}
}

@article{Sarovar2020,
  author  = {Sarovar, M. and Proctor, T. and Rudinger, K. and Young, K. and Nielsen, E. and Blume-Kohout, R.},
  title   = {Detecting crosstalk errors in quantum information processors},
  journal = {Quantum},
  volume  = {4},
  pages   = {321},
  year    = {2020}
}

@article{Krantz2019,
  author  = {Krantz, P. and Kjaergaard, M. and Yan, F. and Orlando, T. P. and Gustavsson, S. and Oliver, W. D.},
  title   = {A quantum engineer's guide to superconducting qubits},
  journal = {Appl. Phys. Rev.},
  volume  = {6},
  pages   = {021318},
  year    = {2019}
}

@article{Chen2014,
  author  = {Chen, Y. and Neill, C. and Roushan, P. and Leung, N. and Fang, M. and Barends, R. and Kelly, J. and Campbell, B. and Chen, Z. and Chiaro, B. and others},
  title   = {Qubit architecture with high coherence and fast tunable coupling},
  journal = {Phys. Rev. Lett.},
  volume  = {113},
  pages   = {220502},
  year    = {2014}
}

@article{Sung2021,
  author  = {Sung, Y. and Ding, L. and Braum{\"u}ller, J. and Veps{\"a}l{\"a}inen, A. and Kannan, B. and Kjaergaard, M. and Greene, A. and Samach, G. O. and McNally, C. and Kim, D. and others},
  title   = {Realization of high-fidelity {CZ} and {ZZ}-free {iSWAP} gates with a tunable coupler},
  journal = {Phys. Rev. X},
  volume  = {11},
  pages   = {021058},
  year    = {2021}
}

@article{Mi2021,
  author  = {Mi, X. and Roushan, P. and Quintana, C. and others},
  title   = {Information scrambling in quantum circuits},
  journal = {Science},
  volume  = {374},
  pages   = {1479},
  year    = {2021}
}

@article{Roushan2017,
  author  = {Roushan, P. and Neill, C. and Tangpanitanon, J. and others},
  title   = {Spectroscopic signatures of localization with interacting photons in superconducting qubits},
  journal = {Science},
  volume  = {358},
  pages   = {1175},
  year    = {2017}
}

@article{Yan2019,
  author  = {Yan, Z. and Zhang, Y.-R. and Gong, M. and others},
  title   = {Strongly correlated quantum walks with a 12-qubit superconducting processor},
  journal = {Science},
  volume  = {364},
  pages   = {753},
  year    = {2019}
}

@article{Jordan1928,
  author  = {Jordan, P. and Wigner, E.},
  title   = {{\"U}ber das {P}aulische {\"A}quivalenzverbot},
  journal = {Z. Phys.},
  volume  = {47},
  pages   = {631},
  year    = {1928}
}

@article{Lieb1961,
  author  = {Lieb, E. and Schultz, T. and Mattis, D.},
  title   = {Two soluble models of an antiferromagnetic chain},
  journal = {Ann. Phys.},
  volume  = {16},
  pages   = {407},
  year    = {1961}
}

@book{Coleman2015,
  author    = {Coleman, P.},
  title     = {Introduction to Many-Body Physics},
  publisher = {Cambridge University Press},
  address   = {Cambridge},
  year      = {2015}
}

@misc{Landi2019,
  author       = {Landi, G. T.},
  title        = {One-dimensional quantum spin chains},
  howpublished = {Lecture notes, University of S{\~a}o Paulo},
  year         = {2019}
}

@article{Peschel2003,
  author  = {Peschel, I.},
  title   = {Calculation of reduced density matrices from correlation functions},
  journal = {J. Phys. A: Math. Gen.},
  volume  = {36},
  pages   = {L205},
  year    = {2003}
}

@article{Prosen2008,
  author  = {Prosen, T.},
  title   = {Third quantization: a general method to solve master equations for quadratic open {F}ermi systems},
  journal = {New J. Phys.},
  volume  = {10},
  pages   = {043026},
  year    = {2008}
}

@article{Zunkovic2014,
  author  = {{\v{Z}}unkovi{\v{c}}, B.},
  title   = {Closed hierarchy of correlations in {M}arkovian open quantum systems},
  journal = {New J. Phys.},
  volume  = {16},
  pages   = {013042},
  year    = {2014}
}

@article{Vidal2004,
  author  = {Vidal, G.},
  title   = {Efficient simulation of one-dimensional quantum many-body systems},
  journal = {Phys. Rev. Lett.},
  volume  = {93},
  pages   = {040502},
  year    = {2004}
}

@book{Breuer2002,
  author    = {Breuer, H.-P. and Petruccione, F.},
  title     = {The Theory of Open Quantum Systems},
  publisher = {Oxford University Press},
  address   = {Oxford},
  year      = {2002}
}

@book{Ring1980,
  author    = {Ring, P. and Schuck, P.},
  title     = {The Nuclear Many-Body Problem},
  publisher = {Springer},
  address   = {Berlin},
  year      = {1980}
}

@article{Lu2021hard,
  author  = {Lu, Lu and Pestourie, Rapha{\"e}l and Yao, Wenjie and Wang, Zhicheng and Verdugo, Francesc and Johnson, Steven G.},
  title   = {Physics-Informed Neural Networks with Hard Constraints for Inverse Design},
  journal = {SIAM J. Sci. Comput.},
  volume  = {43},
  number  = {6},
  pages   = {B1105--B1132},
  year    = {2021},
  doi     = {10.1137/21M1397908}
}

@article{Wu2023,
  author  = {Wu, Chenxi and Zhu, Min and Tan, Qinyang and Kartha, Yadhu and Lu, Lu},
  title   = {A comprehensive study of non-adaptive and residual-based adaptive sampling for physics-informed neural networks},
  journal = {Comput. Methods Appl. Mech. Eng.},
  volume  = {403},
  pages   = {115671},
  year    = {2023},
  doi     = {10.1016/j.cma.2022.115671}
}

@misc{Kingma2015,
  author        = {Kingma, Diederik P. and Ba, Jimmy},
  title         = {Adam: A Method for Stochastic Optimization},
  year          = {2015},
  note          = {ICLR 2015},
  eprint        = {1412.6980},
  archivePrefix = {arXiv}
}

@article{LiuNocedal1989,
  author  = {Liu, Dong C. and Nocedal, Jorge},
  title   = {On the limited memory {BFGS} method for large scale optimization},
  journal = {Math. Program.},
  volume  = {45},
  pages   = {503--528},
  year    = {1989},
  doi     = {10.1007/BF01589116}
}

@inproceedings{Rathore2024,
  author    = {Rathore, Pratik and Lei, Weimu and Frangella, Zachary and Lu, Lu and Udell, Madeleine},
  title     = {Challenges in Training {PINNs}: A Loss Landscape Perspective},
  booktitle = {Proceedings of the 41st International Conference on Machine Learning},
  series    = {PMLR},
  volume    = {235},
  pages     = {42159--42191},
  year      = {2024}
}

\end{document}